\documentclass[twocolumn,showpacs,preprintnumbers,amsmath,amssymb]{revtex4}

\usepackage{graphicx}
\usepackage{dcolumn}
\usepackage{bm}
\usepackage{epsfig}
\usepackage{amssymb}
\usepackage{amsthm}
\usepackage{verbatim}
\usepackage{delarray}
\usepackage{graphics}
\usepackage{epic}

\newcommand{\ket}[1]{{|#1\rangle}}
\newcommand{\bra}[1]{{\langle#1|}}
\newcommand{\braket}[2]{{\langle#1|#2\rangle}}

\newcommand{\bl}{\left(}
\newcommand{\br}{\right)}

\newcommand{\tr}{\mbox{Tr}}

\newcommand{\inner}[2]{\langle{#1},{#2}\rangle}
\newcommand{\deft}{\ {\stackrel{\triangle}{=}} \ }

\newcommand{\hpi}{\widehat{\Pi}}
\newcommand{\hx}{\widehat{X}}
\newcommand{\hd}{\hat{\delta}}

\newcommand{\LL}{{\mathcal{L}}}
\newcommand{\R}{{\mathcal{R}}}
\newcommand{\B}{{\mathcal{B}}}
\newcommand{\N}{{\mathcal{N}}}
\newcommand{\HH}{{\mathcal{H}}}

\newcommand{\C}{{\mathbb{C}}}

\newcommand{\G}{{\mathcal{G}}}
\newcommand{\I}{{\mathcal{I}}}
\newcommand{\Q}{{\mathcal{Q}}}
\newcommand{\SSS}{{\mathcal{S}}}

\newcommand{\ie}{{\em i.e., }}
\newcommand{\eg}{{\em e.g., }}

\newcommand{\etal}{\emph{et al.\ }}

\newtheorem{theorem}{Theorem}

\begin{document}


\title{Mixed quantum state detection with inconclusive results}
\author{Yonina C. Eldar}
 \email{yonina@ee.technion.ac.il}
\affiliation{ Technion---Israel Institute of Technology, Technion
City, Haifa 32000, Israel }

\date{\today}

\begin{abstract}

We consider the problem of designing an optimal quantum detector
with a fixed rate of inconclusive results that maximizes the
probability of correct detection, when distinguishing between a
collection of mixed quantum states. We develop a sufficient
condition for the {\em scaled inverse measurement} to maximize the
probability of correct detection for the case in which the rate of
inconclusive results exceeds a certain threshold. Using this
condition we derive the optimal measurement for linearly
independent pure-state sets, and for mixed-state sets with a broad
class of symmetries. Specifically, we consider geometrically
uniform (GU) state sets and compound geometrically uniform (CGU)
state sets with generators that satisfy a certain constraint.

We then show that the optimal measurements corresponding to GU and
CGU state sets with arbitrary generators are also GU and CGU
respectively, with generators that can be computed very
efficiently in polynomial time within any desired accuracy by
solving a semidefinite programming problem.

\end{abstract}

\maketitle

\section{Introduction}

Quantum information theory refers to the distinctive information
processing properties of quantum systems, which arise when
information is stored in or retrieved from quantum states. A
fundamental aspect of quantum information theory is that
non-orthogonal quantum states cannot be perfectly distinguished.
Therefore, a central problem in quantum mechanics is to design
measurements optimized to distinguish between a collection of
non-orthogonal quantum states.

We consider a quantum state ensemble consisting of $m$ positive
semidefinite Hermitian density operators $\{\rho_i, 1 \le i \le
m\}$ on an $n$-dimensional complex Hilbert space $\HH$, with prior
probabilities $\{p_i>0, 1 \le i \le m\}$. For our {\em
measurement} we  consider general positive operator-valued
measures \cite{P90,P95}, consisting of positive semidefinite
Hermitian operators that form a resolution of the identity on
$\HH$.

Different approaches to distinguishing between the density
operators $\rho_i$ have emerged. In one approach, the measurement
consists of $m$ measurement operators which are designed to
maximize the probability of correct detection. Necessary and
sufficient conditions for an optimum measurement maximizing the
probability of correct detection have been developed
\cite{H73,YKL75,EMV02}. Closed-form analytical expressions for the
optimal measurement have been derived for several special cases
\cite{H76,CBH89,OBH96,BKMO97,EF01,EMV02s}. In particular, the
optimal measurement for pure and mixed-state ensembles with broad
symmetry properties, referred to as geometrically uniform (GU) and
compound GU (CGU) state sets, are considered in
\cite{EF01,EMV02s}. Iterative procedures maximizing the
probability of correct detection have also been developed for
cases in which the optimal measurement cannot be found explicitly
\cite{H82,EMV02}.

More recently, a different approach to the problem has emerged,
which in some cases may be more useful. This approach, referred to
as unambiguous quantum state discrimination, combines  error free
discrimination with a certain fraction of inconclusive results
\cite{I87,D88,P88,JS95,PT98,C98,CB98,E02}. The basic idea,
pioneered by Ivanovic \cite{I87}, is to design a measurement that
with probability $\beta$ returns an inconclusive result, but such
that if the measurement returns an answer, then the answer is
correct with probability $1$. In this case the measurement
consists of $m+1$ measurement operators corresponding to $m+1$
outcomes, where $m$ outcomes correspond to detection of each of
the states and the additional outcome corresponds to an
inconclusive result. Chefles \cite{C98} showed that a necessary
and sufficient condition for the existence of unambiguous
measurements for distinguishing between a collection of quantum
states is that the states are linearly independent pure states.
The optimal measurement minimizing the probability $\beta$ of an
inconclusive result when distinguishing between GU and CGU
pure-state sets was considered in \cite{E02}, and was shown under
certain conditions to be equal to the equal-probability
measurement (EPM).

An interesting alternative approach for distinguishing between a
collection of quantum states, first considered by Chefles and
Barnet \cite{CB982} and Zhang \etal \cite{ZLG99} for pure-state
ensembles, and then later extended by Fiur\'{a}\v{s}ek and
Je\v{z}ek \cite{FJ02} to mixed-state ensembles, is to allow for a
certain probability of an inconclusive result, and then maximize
the probability of correct detection. Thus, in this approach, the
measurement again consists of $m+1$ measurement outcomes; however,
now the outcomes do not necessarily correspond to perfect
detection of each of the states. Indeed, if the quantum states are
mixed states or linearly dependent pure states, then perfect
detection of each of the states is not possible \cite{C98}.
Nonetheless, by allowing for inconclusive results, a higher
probability of correct detection can be obtained in comparison
with the probability of correct detection attainable without
inconclusive results.

Necessary conditions as well as a set of sufficient conditions on
the optimal measurement operators maximizing the probability of
correct detection subject to the constraint that the probability
of an inconclusive result is equal to a constant $\beta$ were
derived in \cite{FJ02}, using Lagrange multiplier theory. It was
also pointed out in \cite{FJ02} that obtaining a closed form
analytical solution to the optimal measurement operators directly
form these conditions is a difficult problem.

In this paper we extend the results of \cite{FJ02} in several
ways. First, using principles of duality  in vector space
optimization, in Section~\ref{sec:conditions} we show that the
conditions derived in \cite{FJ02} are both necessary and
sufficient. We also show that the Lagrange multipliers can be
obtained by solving a reduced size semidefinite programming
problem. This approach lends itself to efficient computational
methods which are guaranteed to converge to the global optimum.

Second, we derive a general condition in Section~\ref{sec:sim}
under which the {\em scaled inverse measurement (SIM)} is optimal.
This measurement consists of measurement operators that are
proportional to the reciprocal states associated with the given
state ensemble, and can be regarded as a generalization of the EPM
to mixed-state ensembles.

Third, we develop the optimal measurement for state sets with
broad symmetry properties. Specifically,
 in Section~\ref{sec:gu} we consider GU
state sets defined over a finite group of unitary matrices. We
obtain a convenient characterization of the SIM and show that the
SIM operators have the same symmetries as the original state set.
We then show that for a pure GU state set and for values of
$\beta$ exceeding a certain threshold, the SIM is optimal. For a
mixed GU state set, under a certain constraint on the generator
and for values of $\beta$ exceeding a threshold, the SIM is again
shown to be optimal.
 For arbitrary values of $\beta$, the optimal
measurement operators corresponding to a pure or mixed GU state
set are shown to be GU with the same generating group, and can be
computed very efficiently in polynomial time.

In Section~\ref{sec:cgu} we consider CGU state sets \cite{EB01},
in which the states are generated by a group of unitary matrices
using {\em multiple} generators.  We obtain a convenient
characterization of the SIM for CGU state sets, and show that the
SIM vectors are themselves CGU. Under a certain condition on the
generators and for values of $\beta$ exceeding a threshold, the
SIM is shown to be optimal. Finally we show that for arbitrary CGU
state sets and for arbitrary values of $\beta$, the optimal
measurement operators are also CGU, and we propose an efficient
algorithm for computing the optimal generators.

It is interesting to note that a closed form analytical expression
exists for the optimal measurement when distinguishing between GU
and CGU (possibly mixed) state sets with generators that satisfy a
certain constraint, under each of the three approaches outlined to
quantum detection, where in the last approach we assume that
$\beta$ exceeds a certain threshold. Furthermore, as shown in
\cite{EMV02s,E02} and in Sections~\ref{sec:gu} and \ref{sec:cgu},
the optimal measurement operators corresponding to GU and CGU
state sets are also GU and CGU respectively, under each one of the
three outlined optimality criteria.

 Before
proceeding to the detailed development, we provide in the next
section a statement of our problem.

\section{Problem Formulation}
\label{sec:problem}

Assume that a quantum channel is prepared in a quantum state drawn
from a collection of given states represented by density operators
$\{ \rho_i,1 \leq i \leq m \}$ on  an $n$-dimensional complex
Hilbert space $\HH$. We assume without loss of generality that the
eigenvectors of $\rho_i,1 \leq i \leq m$, collectively
span\footnote{Otherwise we can transform the problem to a problem
equivalent to the one considered in this paper by reformulating
the problem on the subspace spanned by the eigenvectors of
$\{\rho_i,1 \leq i \leq m\}$. } $\HH$ so that $m \geq n$. Since
$\rho_i$ is Hermitian and positive semidefinite, we can express
$\rho_i$ as $\rho_i=\phi_i\phi_i^*$ for some matrix $\phi_i$, \eg
via the Cholesky or eigendecomposition of $\rho_i$ \cite{GV96}. We
refer to $\phi_i$ as a {\em factor} of $\rho_i$. The choice of
$\phi_i$ is not unique; if $\phi_i$ is a factor of $\rho_i$, then
any matrix of the form $\phi_i'=\phi_iQ_i$ where $Q_i$ is an
arbitrary matrix satisfying $Q_iQ_i^*=I$, is also a factor of
$\rho_i$.

To detect the state of the system a measurement is constructed
comprising $m+1$ measurement operators $\{\Pi_i,0 \leq i \leq m\}$
that satisfy
\begin{eqnarray}
\label{eq:psd} \Pi_i & \geq & 0,\quad 0 \leq i \leq m; \nonumber
\\
\sum_{i=0}^m \Pi_i & = & I.
\end{eqnarray}
Each of the operators $\Pi_i,1 \leq i \leq m$ correspond to
detection of the corresponding states $\rho_i,1 \leq i \leq m$,
and $\Pi_0$ corresponds to an inconclusive result. We seek the
measurement operators $\Pi_i$ that maximize the probability of
correct detection, subject to the constraint that the probability
of an inconclusive result is equal to a constant $\beta<1$.

Given that the transmitted state is $\rho_j$, the probability of
correctly detecting the state using measurement operators
$\{\Pi_i,1 \leq i \leq m\}$ is $\tr(\rho_j\Pi_j)$ and the
probability of a detection error is $\sum_{i=1,i\neq j}^m
\tr(\rho_j\Pi_i)$. Therefore, the probability of correct detection
is given by
\begin{equation}
\label{eq:pe} P_D=\sum_{i=1}^mp_i\tr(\rho_i\Pi_i),
\end{equation}
where $p_i>0$ is the prior probability of $\rho_i$, with $\sum_i
p_i=1$, and the probability of a detection error is given by
\begin{equation}
P_E=\sum_{i=1}^m \sum_{j=1,j \neq i}^m p_i\tr(\rho_i\Pi_j).
\end{equation}
The probability of an inconclusive result is
\begin{equation}
\label{eq:pi} P_I=\sum_{i=1}^m p_i\tr(\rho_i \Pi_0)=\tr(\Delta
\Pi_0)=\beta,
\end{equation}
where for brevity we denote
\begin{equation}
\label{eq:sigma} \Delta=\sum_{i=1}^m p_i\rho_i.
\end{equation}
Our problem is to find the measurement operators $\{\Pi_i,0 \leq i
\leq m\}$ that maximize $P_D$ of (\ref{eq:pe}) subject to the
constraints (\ref{eq:psd}) and (\ref{eq:pi}).

Note that since $\tr(\rho_i)=1$ for all $i$,
\begin{equation}
P_D+P_E+P_I=\sum_{i=1}^m p_i\tr(\rho_i)=1.
\end{equation}
When $P_E=0$ the states are distinguished unambiguously so that if
outcome $i$ is obtained for some $1 \leq i \leq m$, then the state
is $\rho_i$ with probability one. It was shown in \cite{C98} that
with $\beta<1$ we can choose measurement operators such that
$P_E=0$ if and only if the state ensemble is a linearly
independent pure-state ensemble consisting of density operators
$\rho_i$ of the form $\rho_i=\ket{\phi_i}\bra{\phi_i}$ for a set
of linearly independent vectors $\ket{\phi_i}$. If the vectors
$\ket{\phi_i}$ are linearly dependent, or if the ensemble is a
mixed-state ensemble, then $P_E$ cannot be equal to $0$.
Nonetheless, we may seek the measurement operators that minimize
$P_E$, or equivalently, maximize $P_D$, subject to $P_I=\beta$ for
some $\beta<1$.

Equipped with the standard operations of addition and
multiplication by real numbers, the space $\B$ of all Hermitian $n
\times n$ matrices is an $n^2$-dimensional {\em real} vector
space. As noted in \cite{FJ02}, by choosing an appropriate basis
for $\B$, the problem of maximizing $P_D$ subject to
(\ref{eq:psd}) and (\ref{eq:pi}) can be put in the form of a
standard semidefinite programming problem, which is a convex
optimization problem; for a detailed treatment of semidefinite
programming problems see, \eg \cite{A91t,A92,NN94,VB96}. Recently,
methods based on semidefinite programming have been employed in a
variety of different problems in quantum detection and quantum
information \cite{JRF02,DPS02,R01,AM01,EMV02,E02}. By exploiting
the many well known algorithms for solving semidefinite programs
\cite{VB96}, \eg interior point methods\footnote{Interior point
methods are iterative algorithms that terminate once a
pre-specified accuracy has been reached. A worst-case analysis of
interior point methods shows that the effort required to solve a
semidefinite program to a given accuracy grows no faster than a
polynomial of the problem size. In practice, the algorithms behave
much better than predicted by the worst case analysis, and in fact
in many cases the number of iterations is almost constant in the
size of the problem.}  \cite{NN94,A91t}, the optimal measurement
can be computed very efficiently in polynomial time.

 The semidefinite programming formulation can also be used to derive
necessary and sufficient conditions for optimality, which we
discuss in the next section.

\section{Conditions for optimality}
\label{sec:conditions}

Using Lagrange multipliers, it was shown in \cite{FJ02} that a set
of measurement operators $\{\hpi_i,0 \leq i \leq m\}$ maximizes
$P_D$ subject to $P_I=\beta$ for a state set $\{\rho_i,1 \leq i
\leq m\}$ with prior probabilities $\{p_i,1 \leq i \leq m\}$ if
there exists an Hermitian $\hx$ and a constant $\hd$ satisfying
\begin{eqnarray}
\label{eq:condhx1} \hx & \geq & p_i \rho_i,\quad 1 \leq i \leq m;
\\ \label{eq:condhx2}\hx & \geq & \hd \Delta,
\end{eqnarray}
such that
\begin{eqnarray}
\label{eq:condz1} (\hx-p_i\rho_i)\hpi_i&= & 0,\quad 1 \leq i \leq
m; \\
\label{eq:condz2}(\hx-\hd\Delta)\hpi_0 & = & 0.
\end{eqnarray}
It was also shown that (\ref{eq:condz1}) and (\ref{eq:condz2}) are
necessary conditions for optimality.

In Appendix~\ref{app:dual} we use duality arguments similar to
those used in \cite{EMV02} to show that
(\ref{eq:condhx1})--(\ref{eq:condz2}) are {\em necessary and
sufficient} conditions for optimality, so that a set of
measurement operators $\hpi_i$ maximizes $P_D$ subject to
$P_I=\beta$ if and only if there exists an Hermitian $\hx$ and a
constant $\hd$ satisfying (\ref{eq:condhx1})--(\ref{eq:condz2}).
Furthermore, we show that $\hx$ and $\hd$ can be determined as the
solution to the following semidefinite programming problem:
\begin{equation}
\label{eq:min} \min_{X \in \B,\delta \in \R} tr(X)-\delta\beta,
\end{equation}
where $\R$ denotes the reals, subject to
\begin{eqnarray}
\label{eq:condx1} X &\geq &p_i \rho_i,\quad 1 \leq i \leq
m; \nonumber \\
 X &\geq &\delta \Delta.
\end{eqnarray}
The problem of (\ref{eq:min})--(\ref{eq:condx1}) is referred to as
the dual problem.

Note that the dual problem involves many fewer decision variables
than the primal maximization problem. Specifically, in the dual
problem we have $n^2+1$ real decision variables while the primal
problem has $(m+1)n^2$ real decision variables. Therefore, it is
advantageous to solve the dual problem and then use
(\ref{eq:condz1}) and  (\ref{eq:condz2}) to determine the optimal
measurement operators, rather than solving the primal problem
directly.

The necessary conditions (\ref{eq:condhx1}) and (\ref{eq:condz1})
together imply that the rank of each optimal measurement operator
is no larger than the rank of the corresponding density operator;
see \cite{EMV02}. In particular, if the quantum state ensemble is
a pure-state ensemble consisting of (not necessarily independent)
rank-one density operators $\rho_i=\ket{\phi_i}\bra{\phi_i}$, then
the optimal measurement is a pure-state measurement consisting of
rank-one measurement operators $\Pi_i=\ket{\mu_i}\bra{\mu_i}$.

As pointed out in \cite{FJ02}, obtaining a closed-form analytical
expression for the optimal measurement operators directly from the
necessary and sufficient conditions for optimality is a difficult
problem. Since (\ref{eq:min}) is a (convex) semidefinite
programming \cite{VB96,A91t,NN94} problem, there are very
efficient methods for solving (\ref{eq:min}). In particular, the
optimal matrix $\hx$ and optimal scalar $\hd$ minimizing
$\tr(X)-\delta \beta $ subject to (\ref{eq:condx1}) can be
computed in Matlab using the linear matrix inequality (LMI)
Toolbox (see \cite{EMV02,E02} for further details). Once we
determine $\hx$, the optimal measurement operators $\hpi_i$ can be
computed in a similar manner to that described in \cite{EMV02}.

A suboptimal measurement that has been suggested as a detection
measurement for unambiguous quantum state discrimination between
linearly independent pure quantum states, is the EPM
\cite{C98,CB98,E02}, in which the measurement vectors are
proportional to the reciprocal states associated with the states
to be distinguished. A general condition under which the EPM is
optimal for distinguishing between pure quantum states was derived
in \cite{E02}. It was also shown that for GU state sets and for
CGU state sets with generators satisfying a certain constraint,
the EPM is optimal.

In the next section we consider a generalization of the EPM to
mixed quantum states, which we refer to as  the {\em scaled
inverse measurement (SIM}). We then use the necessary and
sufficient conditions for optimality to derive a general condition
under which the SIM is optimal. In Sections~\ref{sec:gu} and
\ref{sec:cgu} we consider some special cases of mixed and
pure-state sets for which the SIM is optimal, and derive explicit
formulas for the optimal measurement operators.

\section{The SIM and the Optimal Measurement}
\label{sec:sim}

The SIM corresponding to a set of density operators
$\{\rho_i=\phi_i\phi_i^*,1 \leq i \leq m\}$ with eigenvectors that
collectively span $\HH$ and prior probabilities $\{p_i,1 \leq i
\leq m\}$ consists of the measurement operators
$\{\Sigma_i=\mu_i\mu_i^*,0 \leq i \leq m\}$ where
\begin{equation}
\label{eq:glsm} \mu_i=\gamma(\Psi\Psi^*)^{-1}\psi_i =\gamma
\Delta^{-1} \psi_i,\quad 1 \leq i \leq m,
\end{equation}
for some $\gamma>0$ and $\Sigma_0=I-\sum_{i=1}^m\mu_i\mu_i^*$.
Here $\Psi$ is the matrix of (block) columns
$\psi_i=\sqrt{p_i}\phi_i$. Note that since the eigenvectors of the
$\{\rho_i\}$ collectively span $\HH$, the columns of the
$\{\psi_i\}$ also together span $\HH$, so $\Psi\Psi^*$ is
invertible. From (\ref{eq:glsm}),
\begin{equation}
\sum_{i=1}^m \mu_i\mu_i^* = \gamma^2\Delta^{-1}\bl \sum_{i=1}^m
\psi_i\psi_i^* \br \Delta^{-1} = \gamma^2\Delta^{-1},
\end{equation}
so that
\begin{equation}
\label{eq:p0}
\Sigma_0=I-\sum_{i=1}^m\mu_i\mu_i^*=I-\gamma^2\Delta^{-1}.
\end{equation}
If follows from (\ref{eq:p0}) that the SIM operators satisfy
(\ref{eq:psd}) if and only if $\gamma^2 \leq \lambda_n$ where
$\{\lambda_i,1 \leq i \leq n\}$ denote the eigenvalues of
$\Delta=\Psi\Psi^*$ and $\lambda_n=\min \lambda_i$.

 In the case in which the prior probabilities
are all equal,
\begin{equation}
\label{eq:lsmep} \mu_i=\gamma (\Phi\Phi^*)^{-1}\phi_i,\quad 1 \leq
i \leq m,
\end{equation}
where $\Phi$ is the matrix of (block) columns $\phi_i$.

Since the factors $\phi_i$ are not unique, the SIM factors $\mu_i$
are also not unique. If $\mu_i$ are the SIM factors corresponding
to $\phi_i$, then the SIM factors corresponding to
$\phi'_i=\phi_iQ_i$ with $Q_iQ_i^*=I$ are $\mu_i'=\mu_iQ_i$.
Therefore, although the SIM factors are not unique, the SIM
operators $\Sigma_i=\mu_i\mu_i^*$ are unique.

The SIM corresponding to a  pure-state ensemble $\ket{\phi_i}$
consists of the measurement vectors $\ket{\mu_i}=\gamma
\Delta^{-1}\ket{\psi_i}$, where
$\ket{\psi_i}=\sqrt{p_i}\ket{\phi_i}$. If in addition
$\gamma=\sqrt{\lambda_n}$, then the SIM vectors are equal to the
EPM vectors \cite{E02}.

The probability of an inconclusive result using the SIM is
\begin{equation}
P_I=\tr(\Delta\Sigma_0)=\tr(\Delta)-\gamma^2\tr(I)=1-n\gamma^2.
\end{equation}
Therefore to satisfy (\ref{eq:pi}),
\begin{equation}
\gamma=\sqrt{\frac{1-\beta}{n}}.
\end{equation}
Since we must also have $\gamma^2 \leq \lambda_n$ we conclude that
$\beta$ must satisfy
\begin{equation}
\label{eq:bmin} \beta \geq 1-n\lambda_n \deft \beta_{\min}.
\end{equation}

For linearly independent pure quantum states it was shown in
\cite{E02} that the SIM with $\gamma=\sqrt{\lambda_n}$ minimizes
the probability of an inconclusive result subject to the
constraint that $P_D=1$ for state sets with strong symmetry
properties. The smallest possible probability of an inconclusive
result in this case is $\beta=\beta_{\min}$. It turns out that for
a large class of state sets, including those discussed in
\cite{E02}, the SIM also maximizes $P_D$ subject to $P_I=\beta$
for $\beta \geq \beta_{\min}$.  From the necessary and sufficient
conditions for optimality discussed in
Section~\ref{sec:conditions} it follows that the SIM is optimal if
and only if the measurement operators $\hpi_i=\mu_i\mu_i^*,1 \leq
i \leq m$ and $\hpi_0=\Sigma_0$ defined by (\ref{eq:glsm}) and
(\ref{eq:p0})
  satisfy (\ref{eq:condz1}) and (\ref{eq:condz2})
for some Hermitian $\hx$ and constant $\hd$ satisfying
(\ref{eq:condhx1}) and (\ref{eq:condhx2}). A sufficient condition
for optimality of the SIM is given in the following theorem, the
proof of which is provided in the Appendix.
\begin{theorem}
\label{thm:gcondition} Let $\{\rho_i=\phi_i\phi_i^*,1 \leq i \leq
m\}$ denote a collection of quantum states with prior
probabilities $\{p_i,1 \leq i \leq m\}$. Let
$\{\Sigma_i=\mu_i\mu_i^*,0 \leq i \leq m\}$ with $\{\mu_i=\gamma
\Delta^{-1}\psi_i,1 \leq i \leq m\}$ and
$\Sigma_0=I-\sum_{i=1}^m\Sigma_i$ denote the scaled inverse
measurement (SIM) operators  corresponding to
$\{\psi_i=\sqrt{p_i}\phi_i,1 \leq i \leq m\}$, where
$\gamma^2=(1-\beta)/n$, $\Delta=\Psi\Psi^*$ and $\Psi$ is the
matrix with block columns $\psi_i$. Let $\lambda_n=\min \lambda_i$
where $\lambda_i$ are the eigenvalues of $\Delta$. Then the SIM
maximizes $P_D$ subject to $P_I=\beta$ for $\beta \geq
\beta_{\min}$ with $\beta_{\min}=1-n\lambda_n$ if for each $1 \leq
i \leq m$,
$(1/\gamma)\mu_i^*\psi_i=\psi_i^*\Delta^{-1}\psi_i=\alpha I$,
where $\alpha$ is a constant independent of $i$.
\end{theorem}

It is interesting to note that the condition of
Theorem~\ref{thm:gcondition} is identical to the condition given
in Theorem~1 of \cite{EMV02s} for the least-squares measurement,
or the square-root measurement, to maximize the probability of
correct detection when $P_I=0$.

As we expect, the condition $\psi_i^*\Delta^{-1}\psi_i=\alpha I$
does not depend on the choice of factor $\phi_i$. Indeed, if
$\phi_i'=\phi_i Q_i$ is another factor of $\rho_i$ with $Q_i$
satisfying $Q_iQ_i^*=I$, and if $\Psi'$ is the matrix of block
columns $\psi_i'=\sqrt{p_i}\phi'_i=\sqrt{p_i}\phi_iQ_i$, then it
is easy to see that $(\psi'_i)^*(\Psi' {\Psi'}^*)^{-1}\psi_i'=\alpha
I$ if and only if $\psi_i^*\Delta^{-1}\psi_i=\alpha I$.

For a pure-state ensemble consisting of density operators
$\rho_i=\ket{\phi_i}\bra{\phi_i}$ for a set of vectors
$\ket{\phi_i}$,  $\braket{\psi_i}{\Delta^{-1}|\psi_i}$ is the
$i$th diagonal element of
$P=\Psi^*\Delta^{-1}\Psi=\Psi^*(\Psi\Psi^*)^{-1}\Psi$. The matrix
$P$ is just the orthogonal projection onto $\N(\Psi)^\perp$, where
$\N(\Psi)$ is the null space of $\Psi$. If the vectors
$\ket{\phi_i}$ are linearly independent, then
$\N(\Psi)^\perp=\{0\}$ so that $P=I$ and
$\braket{\psi_i}{\Delta^{-1}|\psi_i}=1$ for all $i$. It therefore
follows from Theorem~\ref{thm:gcondition} that for a pure-state
ensemble consisting of linearly independent state vectors, the SIM
maximizes $P_D$ subject to $P_I=\beta$ for any $\beta \geq
\beta_{\min}$.


If the state $\rho_i=\phi_i\phi_i^*$ is transmitted with prior
probability $p_i$, then the probability of correctly detecting the
state using measurement operators $\Sigma_i=\mu_i\mu_i^*$ is $p_i
\tr(\mu_i^*\phi_i\phi_i^*\mu_i)= \tr(\mu_i^*\psi_i\psi_i^*\mu_i)$.
It follows that if the condition for optimality of
Theorem~\ref{thm:gcondition} is met, then the probability of
correctly detecting each of the states $\rho_i$ using the SIM is
the same.

For a pure-state ensemble consisting of states $\ket{\phi_i}$ with
prior probabilities $p_i$, the probability of correct detection of
the $i$th state is given by $|\braket{\mu_i}{\psi_i}|^2$.  Since
$\braket{\mu_i}{\psi_i}=\gamma \braket{\psi_i}{\Delta^{-1}|\psi_i}
\geq  0$ for any set of weighted vectors $\ket{\psi_i}$,
$\braket{\mu_i}{\psi_i}$ is constant for all $i$ if and only if
$|\braket{\mu_i}{\psi_i}|^2$ is constant for all $i$.  Therefore,
we may interpret the condition in Theorem~\ref{thm:gcondition} for
pure-state ensembles as follows: The SIM is optimal for a set of
states $\ket{\phi_i}$ with prior probabilities $p_i$ and for
$\beta \geq \beta_{\min}$ if the probability of detecting each one
of the states using the SIM vectors is the same, regardless of the
specific state chosen.

In the remainder of the paper we use Theorem~\ref{thm:gcondition}
to derive the optimal measurement for mixed and (not necessarily
independent) pure-state sets with certain symmetry properties. The
symmetry properties we consider are quite general, and include
many cases of practical interest.

\section{Geometrically Uniform State Sets}
\label{sec:gu}

In this section we  consider {\em geometrically uniform (GU)}
\cite{F91} state sets in which the density operators $\rho_i$ are
defined over a group of unitary matrices and are generated by a
single generating matrix. We first obtain a convenient
characterization of the SIM for GU state sets, and show that under
a certain constraint on the generator the SIM is optimal when
$\beta\geq \beta_{\min}$. In particular, for (not necessarily
independent) pure-state ensembles the SIM is optimal. We then show
that for arbitrary GU state sets and arbitrary values of $\beta$,
the optimal measurement is also GU, and we develop an efficient
computational method for finding the optimal generators.

Let $\G=\{U_i,1 \leq i \leq m\}$ be a finite  group of $m$ unitary
matrices $U_i$. That is, $\G$ contains the identity matrix $I$; if
$\G$ contains $U_i$, then it also contains its inverse $U_i^{-1} =
U_i^*$;  and the product $U_i U_j$ of any two elements of $\G$ is
in $\G$ \cite{A88}.

A state set generated by $\G$ using a single generating operator
$\rho$ is a set $\SSS = \{\rho_i=U_i\rho U_i^*, U_i \in \G\}$. The
group $\G$ is the \emph{generating group} of $\SSS$.  Such a state
set has strong symmetry properties and is called GU. For
consistency with the symmetry of $\SSS$, we will assume
equiprobable prior probabilities on $\SSS$.

If the state set $\{\rho_i,1 \leq i \leq m\}$ is GU, then we can
always choose factors  $\phi_i$ of $\rho_i$ such that
$\{\phi_i=U_i \phi,U_i \in \G\}$ where $\phi$ is a factor of
$\rho$, so that the factors $\phi_i$ are also GU with generator
$\phi$. In the remainder of this section we explicitly assume that
the factors are chosen to be GU.

\subsection{Optimality of the SIM for GU States}

For a GU state set with generating group $\G$, $\Phi\Phi^*$
commutes with each of the matrices $U_i \in \G$
\cite{EB01,EMV02s}. Consequently, $T=(\Phi\Phi^*)^{-1}$ also
commutes with $U_i$ for all $i$, so that from (\ref{eq:lsmep})
\begin{equation}
\label{eq:mui} \mu_i=\gamma T \phi_i=\gamma T U_i\phi=\gamma U_i T
\phi=U_i \mu,\,\, 1 \leq i \leq m,
\end{equation}
where
\begin{equation}
\label{eq:mu} \mu=\gamma(\Phi\Phi^*)^{-1} \phi.
\end{equation}
 It follows that the SIM factors $\mu_i$ are also GU with
generating group $\G$ and generator $\mu$ given by (\ref{eq:mu}).
Therefore, to compute the SIM factors for a GU state set all we
need is to compute the generator $\mu$. The remaining measurement
factors are then obtained by applying the group $\G$ to $\mu$.

From (\ref{eq:mui}) we have that
\begin{equation}
 (1/\gamma)\mu_i^*\psi_i=\frac{1}{\gamma
\sqrt{m}}\mu^*U_i^*U_i\phi= \frac{1}{\gamma\sqrt{m}}\mu^*\phi,
\end{equation}
where $\phi$ and $\mu$ are the generators of the state factors and
the SIM factors, respectively. Thus, the probability of correct
detection of each one of the states $\rho_i$ using the SIM is the
same, regardless of the state transmitted. This then implies from
Theorem~\ref{thm:gcondition} that for a (not necessarily
independent) pure-state GU ensemble the SIM is optimal when $\beta
\geq \beta_{\min}$. For a mixed-state ensemble, if the generator
$\phi$ satisfies
\begin{equation}
\label{eq:gugc}\phi^*(\Phi\Phi^*)^{-1}\phi=\alpha I
\end{equation}
for some $\alpha$, then from Theorem~\ref{thm:gcondition} the SIM
is again optimal.

\subsection{Optimal Measurement for Arbitrary GU States}
\label{sec:agu}

If the generator $\phi$ does not satisfy (\ref{eq:gugc}), or if
$\beta<\beta_{\min}$, then the SIM is no longer guaranteed to be
optimal. Nonetheless, as we now show, the optimal measurement
operators that maximize $P_D$ subject to $P_I=\beta$ for any
$\beta$ are GU with generating group $\G$. The corresponding
generator can be computed very efficiently in polynomial time.

Suppose that the optimal measurement operators  that maximize
\begin{equation}
J(\{\Pi_i\})=\sum_{i=1}^m \tr(\rho_i \Pi_i),
\end{equation}
subject to
\begin{equation}
\label{eq:pip}
 P_I(\{\Pi_i\})=1-\frac{1}{m}\tr\bl\sum_{i,j=1}^m
\rho_i \Pi_j\br=\beta,
\end{equation}
are  $\hpi_i$ and let $\widehat{J}=J(\{\hpi_i\})$. Let $r(j,i)$ be
the mapping from $\I \times \I$ to $\I$ with $\I=\{1,\ldots,m\}$,
defined by $r(j,i)=k$ if $U_j^*U_i=U_k$. Then the measurement
operators $\hpi_i^{(j)}=U_j\hpi_{r(j,i)}U_j^*,1 \leq i \leq m$ and
$\hpi_0^{(j)}=I-\sum_{i=1}^m \hpi_i^{(j)}$ for any $1 \leq j \leq
m$ are also optimal. Indeed, since $\hpi_i \geq 0,1 \leq i \leq m$
and $\sum_{i=1}^m \hpi_i \leq I$, $\hpi^{(j)}_i \geq 0,1 \leq i
\leq m$ and
\begin{equation}
\sum_{i=1}^m \hpi^{(j)}_i=U_j\bl \sum_{i=1}^m \hpi_i \br U_j^*
\leq U_jU_j^*=I.
\end{equation}
Using the fact that $\rho_i=U_i\rho U_i^*$ for some generator
$\rho$,
\begin{eqnarray}
J(\{\hpi^{(j)}_i\})& = & \sum_{i=1}^m\tr(\rho U_i^*
U_j\hpi_{r(j,i)}U_j^*U_i) \nonumber \\
& = & \sum_{k=1}^m\tr(\rho U_k^*\hpi_kU_k) \nonumber \\
& = & \sum_{i=1}^m\tr(\rho_i\hpi_i) \nonumber \\
& = & \widehat{J}.
\end{eqnarray}
Finally,
\begin{eqnarray}
\tr\bl\sum_{i,s=1}^m \rho_i \hpi^{(j)}_s\br & = &
\tr\bl\sum_{i,s=1}^m
U_j^* U_i\rho U_i^* U_j \hpi_{r(j,s)} \br \nonumber \\
& = & \tr\bl\sum_{i,k=1}^m
U_i\rho U_i^* \hpi_k \br \nonumber \\
& = & \tr\bl\sum_{i,k=1}^m \rho_i \hpi_k \br,
\end{eqnarray}
so that from (\ref{eq:pip}),
$P_I(\{\hpi^{(j)}_i\})=P_I(\{\hpi_i\})$.

 Since the measurement operators
$\hpi_i^{(j)}$ are optimal for any $j$, it follows immediately
that the measurement operators
$\{\overline{\Pi}_i=(1/m)\sum_{j=1}^m \hpi_i^{(j)},1 \leq i \leq
m\}$ and $\overline{\Pi}_0=I-\sum_{i=1}^m \overline{\Pi}_i$ are
also optimal. Now, for any $1 \leq i \leq m$,
\begin{eqnarray}
\overline{\Pi}_i & = & \frac{1}{m}\sum_{j=1}^m
U_j\hpi_{r(j,i)}U_j^*
\nonumber \\
 & = & \frac{1}{m}\sum_{k=1}^m U_iU_k^*\hpi_kU_kU_i^* \nonumber \\
 & = & U_i \bl \frac{1}{m}\sum_{k=1}^mU_k^*\hpi_kU_k \br U_i^*
\nonumber \\
 & = & U_i \widehat{\Pi} U_i^*,
\end{eqnarray}
where $\widehat{\Pi}=(1/m)\sum_{k=1}^mU_k^*\hpi_kU_k$.

We therefore conclude that the optimal measurement operators can
always be chosen to be GU with the same generating group $\G$ as
the original state set. Thus, to find the optimal measurement
operators all we need is to find the optimal generator  $\hpi$.
The remaining  operators are obtained by applying the group $\G$
to $\hpi$.

Since the optimal measurement operators satisfy $\Pi_i=U_i \Pi
U_i^*,1 \leq i \leq m$ and $\rho_i=U_i \rho U_i^*$, $\tr(\rho_i
\Pi_i)=\tr (\rho \Pi)$, so that the problem (\ref{eq:pe}) reduces
to the maximization problem
\begin{equation}
\label{eq:max} \max_{\Pi \in \B} \tr(\rho\Pi),
\end{equation}
where $\B$ is the set of $n \times n$ Hermitian operators, subject
to the constraints
\begin{eqnarray}
\label{eq:condp}
&&\Pi  \geq  0; \nonumber \\
&&\sum_{i=1}^m U_i \Pi U_i^* \leq   I; \nonumber \\
&&1- \tr \bl \sum_{i=1}^m U_i \rho U_i \Pi  \br  =  \beta.
\end{eqnarray}
The problem of (\ref{eq:max}) and (\ref{eq:condp}) is a (convex)
semidefinite programming problem, and therefore the optimal $\Pi$
can be computed very efficiently in polynomial time within any
desired accuracy \cite{VB96,A91t,NN94}, for example using the LMI
toolbox on Matlab. Note that the problem of (\ref{eq:max}) and
(\ref{eq:condp}) has $n^2$ real unknowns and $3$ constraints, in
contrast with the original maximization problem (\ref{eq:pe})
subject to (\ref{eq:psd}) and (\ref{eq:pi}) which has $mn^2$ real
unknowns and $m+2$ constraints.

We summarize our results regarding GU state sets in the following
theorem:
\begin{theorem}[GU state sets]
\label{thm:gu} Let $\SSS = \{\rho_i = U_i\rho U_i^*, U_i \in \G\}$
be a geometrically uniform (GU) state set on an $n$-dimensional
Hilbert space, generated by a finite group $\G$ of unitary
matrices, where $\rho=\phi\phi^*$ is an arbitrary generator, and
let $\Phi$ be the matrix of columns $\phi_i=U_i\phi$. Then the
scaled inverse measurement (SIM) is given by the measurement
operators $\Sigma_i=\mu_i\mu_i^*,0 \leq i \leq m$ with
\[\mu_i=U_i\mu,\quad 1 \leq i \leq m,\]
where
\[\mu=\gamma(\Phi\Phi^*)^{-1} \phi,\]
with $\gamma^2=(1-\beta)/n$, and
$\Sigma_0=I-\sum_{i=1}^m\mu_i\mu_i^*$. The SIM has the following
properties:
\begin{enumerate}
\item The measurement operators $\Sigma_i,1 \leq i \leq m$ are GU with generating group $\G$;
\item The probability of correctly detecting each of the states
$\rho_i$ using the SIM is the same;
\item If $\phi^*(\Phi\Phi^*)^{-1} \phi=\alpha I$ for some $\alpha$,
then the SIM maximizes $P_D$ subject to $P_I=\beta$ for $\beta
\geq 1-n\lambda_n$ where $\lambda_n$ is the smallest eigenvalue of
$(1/m)\sum_{i=1}^m \rho_i$ ; In particular, if $\phi=\ket{\phi}$
is a vector so that the state set is a pure-state ensemble, then
the SIM maximizes $P_D$ subject to $P_I=\beta$ for any $\beta \geq
1-n\lambda_n$.
\end{enumerate}
For an arbitrary generator $\phi$ the optimal measurement
operators $\hpi_i,1 \leq i \leq m$ that maximize $P_D$ subject to
$P_I=\beta$ for any $\beta$ are also GU with generating group $\G$
and generator $\Pi$ that maximizes $\tr(\rho \Pi)$ subject to $\Pi
\geq 0, \sum_{i=1}^m U_i \Pi U_i^* \leq I$, and $\tr \bl
\sum_{i=1}^m U_i \rho U_i \Pi \br=1- \beta$.
\end{theorem}

\section{Compound Geometrically Uniform State Sets}
\label{sec:cgu}

We now consider {\em compound geometrically uniform (CGU)}
\cite{EB01} state sets which consist of subsets that are GU. As we
show, the SIM operators are also CGU so that they can be computed
using a {\em set} of generators. Under a certain condition on the
generators and for $\beta\geq \beta_{\min}$, we show that the
optimal measurement associated with a CGU state set is equal to
the SIM. For arbitrary CGU state sets and arbitrary values of
$\beta$  we show that the optimal measurement operators are CGU,
and we derive an efficient computational method for finding the
optimal generators.

A CGU state set is defined as a set of density operators
$\SSS=\{\rho_{ik}=\phi_{ik}\phi_{ik}^*,1 \leq i \leq l,1 \leq k
\leq r\}$ such that $\rho_{ik}=U_i\rho_kU_i^*$, where the matrices
$\{U_i,1 \leq i \leq l\}$ are unitary and form a
 group $\G$, and the operators $\{\rho_k,1 \leq k \leq r\}$  are
the generators. We assume equiprobable prior probabilities on
$\SSS$.

If the state set $\{\rho_{ik},1 \leq i \leq l,1 \leq k \leq r\}$
is CGU, then we can always choose factors  $\phi_{ik}$ of
$\rho_{ik}$ such that $\{\phi_{ik}=U_i \phi_k,1 \leq i \leq l\}$
where $\phi_k$ is a factor of $\rho_k$, so that the factors
$\phi_{ik}$ are also CGU with generators $\{\phi_k,1 \leq k \leq
r\}$. In the remainder of this section we explicitly assume that
the factors are chosen to be CGU.

A CGU state set  is in general not GU. However, for every $k$, the
matrices $\{\phi_{ik},1 \leq i \leq l\}$ and the operators
$\{\rho_{ik},1 \leq i \leq l\}$ are GU with generating group $\G$.
Examples of CGU state sets are considered in \cite{EMV02s}.

\subsection{Optimality of the SIM for CGU State Sets}

With $\Phi$ denoting the matrix of (block) columns $\phi_{ik}$, it
was shown in \cite{EB01,EMV02s} that $\Phi\Phi^*$, and
consequently $T=(\Phi\Phi^*)^{-1}$, commutes with each of the
matrices $U_i \in \G$. Thus, the  SIM operators are
$\Sigma_{ik}=\mu_{ik}\mu_{ik}^*,1 \leq i \leq l,1 \leq k \leq r$
with
\begin{equation}
\label{eq:muik} \mu_{ik}=\gamma T \phi_{ik}=\gamma T U_i\phi_k=U_i
\mu_k,
\end{equation}
where
\begin{equation}
\label{eq:muc} \mu_k=\gamma T \phi_k=\gamma (\Phi\Phi^*)^{-1}
\phi_k.
\end{equation}
Therefore the SIM factors are also CGU with generating group $\G$
and generators $\mu_k$ given by (\ref{eq:muc}). To compute the SIM
factors all we need is to compute  the generators $\mu_k$. The
remaining measurement factors are then obtained by applying the
group $\G$ to each of the generators.

From (\ref{eq:muik}),
\begin{equation}
\label{eq:cgu_condt}
\mu_{ik}^*\phi_{ik}=\mu_k^*U_i^*U_i\phi_{k}=\mu_k^*\phi_{k},
\end{equation}
so that from Theorem~\ref{thm:gcondition} the SIM is optimal if
\begin{equation}
\label{eq:cgu_cond} \mu^*_k\phi_k=\gamma \phi_k^*T\phi_k=\alpha
I,\quad 1 \leq k \leq r,
\end{equation}
for some constant $\alpha$.

\subsection{CGU State Sets  With GU Generators}
\label{sec:cgugu}

A special class of CGU state sets  is {\em CGU state sets  with GU
generators} in which the generators $\{\rho_k=\phi_k\phi_k^*,1
\leq k \leq r\}$ and the factors $\phi_k$ are themselves GU.
Specifically, $\{\phi_k=V_k \phi\}$ for some generator $\phi$,
where the matrices $\{V_k,1 \leq k \leq r\}$ are unitary, and form
a group $\Q$.

Suppose that $U_i$ and $V_k$ commute up to a phase factor for all
$i$ and $k$ so that $U_iV_k=V_kU_ie^{j\theta(i,k)}$ where
$\theta(i,k)$ is an arbitrary phase function that may depend on
the indices $i$ and $k$. In this case we say that $\G$ and $\Q$
commute up to a phase factor and that the corresponding state set
is {\em CGU with commuting GU generators}. (In the  special case
in which $\theta=0$ so that $U_iV_k=V_kU_i$ for all $i,k$, the
resulting state set is GU \cite{EB01}). Then for  all $i,k$,
$\Phi\Phi^*$ commutes with $U_iV_k$ \cite{EMV02s}, and the SIM
factors $\mu_{ik}$ are given by
\begin{equation}
\label{eq:LSMcgugu} \mu_{ik}= \gamma T\phi_{ik}=\gamma T U_iV_k
\phi=U_i V_k\bar{\mu},
\end{equation}
where $\bar{\mu}=\gamma T \phi$. Thus even though the state set is
not in general GU, the SIM factors can be computed using a single
generator.

Alternatively, we can express $\mu_{ik}$ as $\mu_{ik}=U_i\mu_k$
where the generators $\mu_k$ are given by
\begin{equation}
\label{eq:LSMg} \mu_k=V_k\bar{\mu}.
\end{equation}
From (\ref{eq:LSMg}) it follows that the generators $\mu_k$ are GU
with generating group $\Q=\{V_k,1 \leq k \leq r\}$ and generator
$\bar{\mu}$. Then  for all $k$,
\begin{equation}
\label{eq:gc} \mu_k^*\phi_k= \bar{\mu}^*V_k^*V_k\phi=
\bar{\mu}^*\phi.
\end{equation}
If in addition,
\begin{equation}
\label{eq:cmguc} \bar{\mu}^*\phi=\gamma\phi^*T\phi=\alpha I
\end{equation}
for some $\alpha$, then combining (\ref{eq:cgu_condt}),
(\ref{eq:gc}) and (\ref{eq:cmguc}) with
Theorem~\ref{thm:gcondition} we conclude that the SIM is optimal.
In particular, for a pure-state ensemble, $\bar{\mu}^*\phi$ is a
scalar so that (\ref{eq:cmguc}) is always satisfied. Therefore,
for a pure CGU state set with commuting GU generators, the SIM
maximizes $P_D$ subject to $P_I=\beta$ for $\beta \geq
\beta_{\min}$.

\subsection{Optimal Measurement for Arbitrary CGU States}

If the generators $\phi_k$ do not satisfy (\ref{eq:cgu_cond}), or
if $\beta<\beta_{\min}$, then the SIM is no longer guaranteed to
be optimal. Nonetheless, as we now show, the optimal measurement
operators that maximize $P_D$ subject to $P_I=\beta$ are CGU with
generating group $\G$. The corresponding generators can be
computed very efficiently in polynomial time within any desired
accuracy.

Suppose that the optimal measurement operators  that maximize
\begin{equation}
J(\{\Pi_{ik}\})=\sum_{i=1}^l \sum_{k=1}^r\tr(\rho_{ik} \Pi_{ik}),
\end{equation}
subject to
\begin{equation}
\label{eq:pipc}
 P_I(\{\Pi_{ik}\})=1-\frac{1}{lr}\tr\bl\sum_{i,j=1}^l \sum_{k,s=1}^r
\rho_{ik} \Pi_{js} \br=\beta,
\end{equation}
are $\hpi_{ik}$, and let $\widehat{J}=J(\{\hpi_{ik}\})$. Let
$r(j,i)$ be the mapping from $\I \times \I$ to $\I$ with
$\I=\{1,\ldots,l\}$, defined by $r(j,i)=s$ if $U_j^*U_i=U_s$. Then
the measurement operators $\hpi_{ik}^{(j)}=U_j\hpi_{r(j,i)k}U_j^*$
for any $1 \leq j \leq l$ are also optimal. Indeed, since
$\hpi_{ik} \geq 0,1 \leq i \leq l,1 \leq k \leq r$ and
$\sum_{i=1}^l\sum_{k=1}^r \hpi_{ik} \leq I$, $\hpi^{(j)}_{ik} \geq
0,1 \leq i \leq l,1 \leq k \leq r$ and
\begin{equation}
\sum_{i=1}^l \sum_{k=1}^r \hpi^{(j)}_{ik}=U_j\bl \sum_{i=1}^l
\sum_{k=1}^r \hpi_{ik} \br U_j^* \leq U_jU_j^*=I.
\end{equation}
Using the fact that $\rho_{ik}=U_i\rho_k U_i^*$ for some
generators $\rho_k$,
\begin{eqnarray}
J(\{\hpi^{(j)}_{ik}\}) & = & \sum_{i=1}^l \sum_{k=1}^r \tr(\rho_k
U_i^* U_j\hpi_{r(j,i)k}U_j^*U_i) \nonumber \\
& = & \sum_{s=1}^l \sum_{k=1}^r\tr(\rho_k U_s^*\hpi_{sk}U_s)
\nonumber \\ & = & \sum_{i=1}^l \sum_{k=1}^r\tr(\rho_{ik}
\hpi_{ik}) \nonumber \\& = &  \widehat{J}.
\end{eqnarray}
Finally,
\begin{eqnarray}
\lefteqn{ \tr\bl\sum_{i,s=1}^l \sum_{k,t=1}^r \rho_{ik}
\hpi^{(j)}_{st}\br=} \nonumber \\
 & = & \tr\bl \sum_{i,s=1}^l
\sum_{k,t=1}^r
U_j^* U_i\rho_k U_i^* U_j \hpi_{st} \br \nonumber \\
& = & \tr\bl \sum_{i,s=1}^l \sum_{k,t=1}^r
U_i\rho_k U_i^*  \hpi_{st} \br \nonumber \\
& = & \tr\bl \sum_{i,s=1}^l \sum_{k,t=1}^r \rho_{ik} \hpi_{st}
\br,
\end{eqnarray}
so that from (\ref{eq:pipc}),
$P_I(\{\hpi^{(j)}_{ik}\})=P_I(\{\hpi_{ik}\})$.

Since the measurement operators $\hpi^{(j)}_{ik}$ are optimal for
any $j$, it follows immediately that the measurement operators
$\{\overline{\Pi}_{ik}=(1/l)\sum_{j=1}^l \hpi^{(j)}_{ik},1 \leq i
\leq l,1 \leq k \leq r\}$ and $\overline{\Pi}_{0}=I-\sum_{i,k}
\overline{\Pi}_{ik}$
 are also optimal.
Now, for any $1 \leq i \leq l, 1 \leq k \leq r$,
\begin{eqnarray}
\overline{\Pi}_{ik} & = & \frac{1}{l}\sum_{j=1}^l
U_j\hpi_{r(j,i)k}U_j^*
\nonumber \\
 & = & \frac{1}{l}\sum_{s=1}^l U_iU_s^*\hpi_{sk}U_sU_i^* \nonumber \\
 & = & U_i \bl \frac{1}{l}\sum_{s=1}^lU_s^*\hpi_{sk}U_s \br U_i^*
\nonumber \\
 & = & U_i \widehat{\Pi}_k U_i^*,
\end{eqnarray}
where $\widehat{\Pi}_k=(1/l)\sum_{s=1}^lU_s^*\hpi_{sk}U_s$.

We therefore conclude that the optimal measurement operators can
always be chosen to be CGU with the same generating group $\G$ as
the original state set. Thus, to find the optimal measurement
operators all we need is to find the optimal generators
$\{\hpi_k,1 \leq k \leq r\}$ . The remaining operators are
obtained by applying the group $\G$ to each of the generators.

Since the optimal measurement operators satisfy $\Pi_{ik}=U_i
\Pi_k U_i^*$ and $\rho_{ik}=U_i \rho_k U_i^*$, $\tr(\rho_{ik}
\Pi_{ik})=\tr (\rho_k \Pi_k)$, so that the problem (\ref{eq:pe})
reduces to the maximization problem
\begin{equation}
\label{eq:max2} \max_{\Pi_k \in \B} \sum_{k=1}^r\tr(\rho_k\Pi_k),
\end{equation}
subject to the constraints
\begin{eqnarray}
\label{eq:condp2}
&& \Pi_k  \geq  , \quad 1 \leq k \leq r; \nonumber \\
&&\sum_{i=1}^l\sum_{k=1}^r  U_i \Pi_k U_i^*  \leq  I; \nonumber \\
&& 1- \frac{1}{r}\tr \bl \sum_{i=1}^l\sum_{k,l=1}^r U_{i} \rho_k
U_{i} \Pi_l \br  = \beta.
\end{eqnarray}
Since this problem is a (convex) semidefinite programming problem,
the optimal generators $\Pi_k$ can be computed very efficiently in
polynomial time within any desired accuracy \cite{VB96,A91t,NN94},
for example using the LMI toolbox on Matlab. Note that the problem
of (\ref{eq:max2}) and (\ref{eq:condp2}) has $rn^2$ real unknowns
and $r+2$ constraints, in contrast with the original maximization
(\ref{eq:pe}) subject to (\ref{eq:psd}) and (\ref{eq:pi}) which
has $lrn^2$ real unknowns and $lr+2$ constraints.

We summarize our results regarding CGU state sets in the following
theorem:
\begin{theorem}[CGU state sets]
\label{thm:cgu} Let $\SSS = \{\rho_{ik} = U_i\rho_k U_i^*, 1 \leq
i \leq l,1 \leq k \leq r\}$ be a compound geometrically uniform
(CGU) state set on an $n$-dimensional Hilbert space generated by a
finite group $\G$ of unitary matrices and generators
$\{\rho_k=\phi_k\phi_k^*,1 \leq k \leq r\}$, and let $\Phi$ be the
matrix of columns $\phi_{ik}=U_i\phi_k$. Then the scaled inverse
measurement (SIM) is given by the measurement operators
$\Sigma_{ik}=\mu_{ik}\mu_{ik}^*,1 \leq i \leq l,1 \leq k \leq r$
and $\Sigma_0=I-\sum_{i,k}\mu_{ik}\mu_{ik}^*$ with
\[\mu_{ik}=U_i\mu_k\]
where
\[\mu_k=\gamma (\Phi\Phi^*)^{-1} \phi_k,\]
and $\gamma^2=(1-\beta)/n$.
 The SIM has the following properties:
\begin{enumerate}
\item The measurement operators $\Sigma_{ik},1 \leq i \leq l,1 \leq k \leq r$
are CGU with generating group $\G$;
\item The probability of correctly detecting each of the states
$\phi_{ik}$ for fixed $k$ using the SIM is the same;
\item If  $\phi_k^*(\Phi\Phi^*)^{-1} \phi_k=\alpha I$ for some
$\alpha$ and for $1 \leq k \leq r$, then the SIM maximizes $P_D$
subject to $P_I=\beta$ with $\beta \geq 1-n\lambda_n$ where
$\lambda_n$ is the smallest eigenvalue of $(1/lr)\sum_{i,k}
\rho_{ik}$.
\end{enumerate}
If in addition the generators $\{\phi_k=V_k\phi,1 \leq k \leq r\}$
are geometrically uniform with $U_iV_k=V_kU_ie^{j\theta(i,k)}$ for
all $i,k$, then
\begin{enumerate}
\item $\mu_{ik}=U_iV_k\bar{\mu}$ where
$\bar{\mu}=\gamma(\Phi\Phi^*)^{-1}\phi$ so that the SIM operators
are CGU with geometrically uniform generators;
\item The probability of correctly detecting each of the states
$\phi_{ik}$ using the SIM is the same;
\item If  $\phi^*(\Phi\Phi^*)^{-1} \phi=\alpha I$ for some
$\alpha$,
 then the SIM
maximizes $P_D$ subject to $P_I=\beta$ with $\beta \geq
1-n\lambda_n$. In particular, if $\phi=\ket{\phi}$ is a vector so
that the state set is a pure-state ensemble, then the SIM
maximizes $P_D$ subject to $P_I=\beta$ with $\beta \geq
1-n\lambda_n$.
\end{enumerate}
For arbitrary CGU state sets the optimal measurement operators
$\hpi_{ik},1 \leq i \leq l,1 \leq k \leq r$ that maximize $P_D$
subject to $P_I=\beta$ for any $\beta$ are CGU with generating
group $\G$ and generators $\Pi_k$ that maximize $\sum_{k=1}^r
\tr(\rho_k \Pi_k)$ subject to $\Pi_k \geq 0,1 \leq k \leq r$,
$\sum_{i,k} U_i \Pi_k U_i^* \leq I$, and $\tr \bl
\sum_{i=1}^l\sum_{k,l=1}^r U_{i} \rho_k U_{i} \Pi_l  \br =
r(1-\beta)$.
\end{theorem}

\section{Conclusion}

In this paper we considered the optimal measurement operators that
maximize the probability of correct detection given a fixed
probability $\beta$ of an inconclusive result, when distinguishing
between a collection of {\em mixed} quantum states. We first
derived a set of necessary and sufficient conditions for
optimality by exploiting principles of duality theory in vector
space optimization. Using these conditions, we derived a general
condition under which the SIM is optimal. We then considered state
sets with a broad class of symmetry properties for which the SIM
is optimal. Specifically, we showed that for GU state sets and for
CGU state sets with generators that satisfy certain constraints
and for values of $\beta$ exceeding a threshold,  the SIM is
optimal. We also showed that for arbitrary GU and CGU state sets
and for arbitrary values of $\beta$, the optimal measurement
operators have the same symmetries as the original state sets.
Therefore, to compute the optimal measurement operators, we need
only to compute the corresponding generators. As we showed, the
generators can be computed very efficiently in polynomial time
within any desired accuracy by solving a semidefinite programming
problem.

\begin{acknowledgments}

The author wishes to thank Prof. A.\ Megretski and Prof.\ G.\ C.\
Verghese for valuable discussions that lead to many of the results
in this paper.

\end{acknowledgments}

\appendix

\section{Necessary and Sufficient conditions for optimality}
\label{app:dual}

Denote by $\Lambda$  the set of all ordered sets
$\Pi=\{\Pi_i\}_{i=0}^m, \Pi_i \in \B$ satisfying (\ref{eq:psd})
and (\ref{eq:pi}) with $\beta<1$, and define $J(\Pi)= \sum_{i=1}^m
p_i\tr(\rho_i\Pi_i)$. Then our problem is
\begin{equation}
\label{eq:primal} \max_{\Pi \in \Lambda} J(\Pi).
\end{equation}
We refer to this problem as the primal problem, and to any $\Pi
\in \Lambda$ as a primal feasible point. The optimal value of
$J(\Pi)$ is denoted by $\widehat{J}$.

To derive necessary and sufficient conditions for optimality, we
now formulate a {\em dual problem} whose optimal value serves as a
certificate for $\widehat{J}$. As described in \cite{EMV02}, a
general method for deriving a dual problem is to invoke the
separating hyperplane theorem \cite{L68}, which states that two
disjoint convex sets\footnote{A set $C$ is convex if for any $x,y
\in C$, $\alpha x+(1-\alpha)y \in C$ for all $\alpha \in [0,1]$.}
can always be separated by a hyperplane. We will take one convex
set to be the point $0$, and then carefully construct another
convex set that does not contain $0$, and that captures the
 equality constraints in the primal problem and the fact
that for any primal feasible point, the value of the primal
function is no larger than the optimal value. The dual variables
will then emerge from the parameters of the separating hyperplane.

In our problem we have two equality constraints,
$\sum_{i=0}^m\Pi_i=I$ and $\tr(\Delta\Pi_0)=\beta$ and we know
that $\widehat{J} \geq J(\Pi)$. Our constructed convex set will
accordingly consist of matrices of the form $-I+\sum_{i=0}^m\Pi_i$
where $\Pi_i \in \B$ and $\Pi_i \geq 0$, scalars of the form
$\beta-\tr(\Delta\Pi_0)$, and scalars of the form $r -J(\Pi)$
where $r>\widehat{J}$. We thus consider the real vector space
\[\LL=\B\times\R\times \R=\{(S,x,y):\ \ S\in \B,\ x,y\in\R\},\]
where $\R$ denotes the reals, with inner product defined by
\begin{equation}
 \inner{(W,z,t)}{(S,x,y)}=\tr(WS)+zx+ty.
\end{equation}
We then define the subset $\Omega$  of $\LL$ as points of the form
\begin{equation}
\Omega=\left(-I+\sum_{i=0}^m\Pi_i, \beta-\tr\bl \Delta \Pi_0\br,
r-\sum_{i=1}^mp_i\tr(\Pi_i\rho_i)\right),
\end{equation}
where $ \Pi_i\in \B, \Pi_i\ge0, r\in\R$  and $r>\widehat{J}$.

It is easily verified that $\Omega$ is convex, and
$0\not\in\Omega$. Therefore, by the separating hyperplane theorem,
there exists a nonzero vector $(Z,a,b) \in \LL$ such that
$\inner{(Z,a,b)}{(Q,c,d)} \geq 0$ for all $(Q,c,d)\in\Omega$, \ie
\begin{eqnarray}
\label{eq:hyperplane} \lefteqn{\hspace*{-0.6in}
\tr\left(Z\left(-I+\sum_{i=0}^m\Pi_i\right)\right)+
b\bl\beta-\tr\bl\Delta \Pi_0\br \br+} \nonumber \\
&&+ a\left(r-\sum_{i=1}^mp_i\tr(\Pi_i\rho_i)\right)\ge0
\end{eqnarray}
for all $\Pi_i\in \B$ and $r\in\R$ such that $\Pi_i\ge0$,
$r>\widehat{J}$.

As we now show, the hyperplane parameters $(Z,a,b)$ have to
satisfy certain constraints, which lead to the formulation of the
dual problem. Specifically, (\ref{eq:hyperplane}) with $\Pi_i=0$,
$r\to \widehat{J}$ implies
\begin{equation}
\label{eq:a} a\widehat{J}\ge\tr(Z)-b\beta.
\end{equation}
Similarly, (\ref{eq:hyperplane}) with $r=\widehat{J}+1$, $\Pi_j=0$
for $j\neq i$, $\Pi_i=t\ket{x}\bra{x}$ for one value $1 \leq i
\leq m$ where $\ket{x}\in\C^n$ is fixed and $t\to+\infty$ yields
$\braket{x}{Z-ap_i\rho_i|x}\ge 0$. Since $\ket{x}$ and $i$ are
arbitrary, this implies
\begin{equation}
\label{eq:Z} Z\ge ap_i\rho_i,\quad 1 \leq i \leq m.
\end{equation}
With $r=\widehat{J}+1$, $\Pi_j=0$ for $j\neq 0$,
$\Pi_0=t\ket{x}\bra{x}$ where $\ket{x}\in\C^n$ is fixed and
$t\to+\infty$, (\ref{eq:hyperplane})
 yields $\braket{x}{Z-b\Delta|x}\ge 0$, which implies
\begin{equation}
\label{eq:Z2} Z\ge b\Delta.
\end{equation}

With $\Pi_i=0,0 \leq i \leq m$, $r\to+\infty$,
(\ref{eq:hyperplane}) implies $a\ge0$. If  $a=0$, then
(\ref{eq:a}) yields $\tr(Z) \leq b\beta<b$ and (\ref{eq:Z2})
yields $\tr(Z) \geq b$. Therefore we conclude that $a>0$, and
define $\hx=Z/a$, $\hd=b/a$. Then (\ref{eq:a}) implies that
\begin{equation}
\label{eq:tx} T(\hx,\hd)\le \widehat{J},
\end{equation}
where $T(X,\delta)=\tr(X)-\delta\beta$, (\ref{eq:Z}) implies that
$\hx \geq p_i\rho_i$ for $1 \leq i \leq m$, and (\ref{eq:Z2})
implies that $\hx \geq \hd \Delta$.

Let $\Gamma$ be the set of $X \in \B$, $\delta \in \R$ satisfying
$X \geq p_i\rho_i,1 \leq i \leq m$ and $X \geq \delta \Delta$.
 Then for any
 $X,\delta\in\Gamma$,
$\Pi \in\Lambda$, we have
\begin{eqnarray}
\label{eq:wduality} \lefteqn{\hspace*{-0.5in}
T(X,\delta)-J(\Pi)=\tr\bl\sum_{i=1}^m\Pi_i(X-p_i\rho_i)\br} \nonumber \\
&&+\tr \bl \Pi_0(X-\delta\Delta) \br \ge 0.
\end{eqnarray}
Since $\hx \in \Gamma$, from (\ref{eq:tx}) and (\ref{eq:wduality})
we conclude that $T(\hx,\hd)=\widehat{J}$.

Thus we have proven that the dual problem associated with
(\ref{eq:primal}) is
\begin{equation}
\label{eq:min2} \min_{X \in \B,\delta \in \R} tr(X)-\delta\beta,
\end{equation}
subject to
\begin{eqnarray}
 X &\geq &p_i \rho_i,\quad 1 \leq i \leq
m\nonumber; \\
X &\geq &\delta \Delta.
\end{eqnarray}
Furthermore, we have shown that there exists  an optimal $\hx,\hd
\in\Gamma$ and an optimal value $\widehat{T}=T(\widehat{X},\hd)$
such that
 $\widehat{T}=\widehat{J}$.

 Let $\hpi_i$ denote the optimal measurement operators. Then
combining (\ref{eq:wduality}) with  $\widehat{T}=\widehat{J}$, we
conclude that
\begin{eqnarray}
\label{eq:condz1a} (\hx-p_i\rho_i)\hpi_i&= & 0,\quad 1 \leq i \leq
m; \nonumber \\
(\hx-\hd\Delta)\hpi_0 & = & 0.
\end{eqnarray}
Once we find the optimal $\hx$ and $\hd$ that minimize the dual
problem (\ref{eq:min2}), the constraints (\ref{eq:condz1a})
 are necessary and sufficient conditions on the
optimal measurement operators $\hpi_i$. We have already seen that
these conditions are necessary. To show that they are sufficient,
we note that if a set of feasible measurement operators $\Pi_i$
satisfies (\ref{eq:condz1a}), then $\sum_{i=1}^m\tr \bl
\Pi_i(\hx-p_i\rho_i)\br= 0$ and $\tr \bl (\hx-\hd\Delta)\hpi_0
\br= 0$ so that from (\ref{eq:wduality}),
$J(\Pi)=T(\hx,\hd)=\widehat{J}$.

\section{Proof of Theorem~\ref{thm:gcondition}}

In this appendix we prove Theorem~\ref{thm:gcondition}.
Specifically, we show that for a set of states
$\rho_i=\phi_i\phi_i^*$ with prior probabilities $p_i$, if
$(1/\gamma)\mu_i^*\psi_i=\alpha I,1 \leq i \leq m$, where
$\mu_i=\gamma (\Psi\Psi^*)^{-1}\psi_i=\gamma \Delta^{-1}\psi_i$
are the SIM factors and $\psi_i=\sqrt{p_i} \phi_i$, then there
exists an Hermitian $X$ and a constant $\delta$ such that
\begin{eqnarray}
\label{eq:appc1}
&&X  \geq \psi_i\psi_i^*,\quad 1 \leq i \leq m; \\
\label{eq:appc2}
&&X \geq  \delta \Delta; \\
\label{eq:appc3} &&(X-\psi_i\psi_i^*)\mu_i\mu_i^*  =  0,\quad 1
\leq i \leq m;  \\
&&\label{eq:appc4} (X-\delta \Delta)(I-\gamma^2 \Delta^{-1})  =
0.
\end{eqnarray}

Let $X=\alpha \Delta$ and $\delta=\alpha$. Then (\ref{eq:appc2})
and (\ref{eq:appc4}) are immediately satisfied. Next, since
$\alpha
I=\psi_i^*\Delta^{-1}\psi_i=\psi_i^*\Delta^{-1/2}\Delta^{-1/2}\psi_i$,
it follows that
\begin{equation}
\label{eq:tmp} \alpha I \geq
\Delta^{-1/2}\psi_i\psi_i^*\Delta^{-1/2}.
\end{equation}
Multiplying both sides of (\ref{eq:tmp}) by $\Delta^{1/2}$ we have
\begin{equation}
\alpha \Delta \geq \psi_i\psi_i^*,
\end{equation}
which verifies that the conditions (\ref{eq:appc1}) are satisfied.

Finally,
\begin{equation}
(X-\psi_i\psi_i^*)\mu_i= \alpha\gamma
\Delta\Delta^{-1}\psi_i-\alpha \gamma \psi_i=0,
\end{equation}
so that the conditions (\ref{eq:appc3}) are also satisfied.


\end{document}